\documentclass[namedreferences]{SolarPhysics}
\usepackage[optionalrh]{spr-sola-addons} % For Solar Physics 
\usepackage{epsfig}                     % For eps figures, old commands
\usepackage{graphicx}                    % For eps figures, newer & more powerfull
\usepackage{color}                       % For color text: \color command
\usepackage{url}                         % For breaking URLs easily trough lines
\begin{document}
\begin{article}
\begin{opening}
\title{Solar Cycle Characteristics Examined in Separate Hemispheres:
    Phase, Gnevyshev Gap, and Length of Minimum}

%%%%%%%%%%%%%%%%%%%%%%%%%%%%%%%%%%%%%%%%%%%%%%%%%%%
%% Authors Names
\author{A.A.~\surname{Norton}$^{1,2}$\sep
        J.C.~\surname{Gallagher}$^{1}$      
       }

%%%%%%%%%%%%%%%%%%%%%%%%%%%%%%%%%%%%%%%%%%%%%%%%%%%
%% Runningheads
\runningauthor{Norton \& Gallagher}
\runningtitle{Solar Cycle Characteristics Examined in Separate Hemispheres}

%%%%%%%%%%%%%%%%%%%%%%%%%%%%%%%%%%%%%%%%%%%%%%%%%%%
%% Affiliations 
  \institute{$^{1}$ National Solar Observatory, 950 N. Cherry Ave., Tucson, AZ 85719, USA\\
             $^{2}$ Centre for Astronomy, James Cook University, Townsville, QLD 4811, Australia
                     email: \url{Aimee.Norton@JCU.edu.au}\\
             }

%%%%%%%%%%%%%%%%%%%%%%%%%%%%%%%%%%%%%%%%%%%%%%%%%%%
%%% Abstract 
\begin{abstract}
According to recent research results from solar dynamo models, the north and south hemispheres may evolve separately throughout the solar cycle. The observed phase lag between the north and south hemispheres provides information regarding how strongly the hemispheres are coupled. Using hemispheric sunspot-area and sunspot-number data from cycles $12-23$, we determine how out of phase the separate hemispheres are during the rising, maximum, and declining period of each solar cycle.  Hemispheric phase differences range from $0-11$, $0-14$, and $2-19$ months for the rising, maximum, and declining periods, respectively. The phases appear randomly distributed between 0 months (in phase) and half of the rise (or decline) time of the solar cycle.  An analysis of the sunspot cycle double peak, or Gnevyshev gap, is conducted to determine 
if the double-peak is caused by the averaging of two hemispheres that are out of phase. We confirm previous findings that the Gnevyshev gap is a phenomena that occurs in the separate hemispheres and is not due to a superposition of sunspot indices from hemispheres slightly out of phase. Cross hemispheric coupling could be strongest at solar minimum, when there are large quantities of magnetic flux at the Equator. We search for a correlation between the hemispheric phase difference near the end of the solar cycle and the length of solar cycle minimum, but found none. Because magnetic flux diffusion across the Equator is a mechanism by which the hemispheres couple, we measured the magnetic flux crossing the Equator by examining Kitt Peak Vacuum Telescope and SOLIS magnetograms for solar cycles $21-23$.  We find, on average, a surplus of northern hemisphere magnetic flux crossing during the mid-declining phase of each solar cycle. However, we find no correlation between magnitude of magnetic flux crossing the Equator, length of solar minima, and phase lag between the hemispheres.
\end{abstract}

%%%%%%%%%%%%%%%%%%%%%%%%%%%%%%%%%%%%%%%%%%%%%%%%%%%
%% Keywords
\keywords{Solar Cycle, Observations; Sunspots, Statistics}

\end{opening}
%-------------------------------------------------

%%%%%%%%%%%%%%%%%%%%%%%%%%%%%%%%%%%%%%%%%%%%%%%%%%%
%% Sections

\section{Introduction}

Many aspects of the solar cycle remain unknown, including the strength and the mechanism by which the north and south hemispheres couple.  The most accepted dynamo model is the Babcock--Lieghton model, which has as its main constituents the creation of toroidal fields from a shearing of the poloidal field by differential rotation within the solar interior, and the creation of poloidal field by the decay of bipolar active regions on the solar surface (Babcock, 1961; Leighton, 1964, 1969).  The proposed Babcock--Leighton mechanism of generating poloidal fields, based on observations of erupting bipolar magnetic regions in the presence of supergranular diffusion, can be tested and modeled in a detailed fashion due to the ability to observe the solar surface.  The modeling efforts of Wang and Sheeley (1991) confirmed that the proposed mechanism was robust even in the presence of observed rates of decay, supergranular diffusion, meridional flows and including the effects of ephemeral regions. 

The Babcock--Leighton dynamo models most widely used to reproduce the solar cycle assume a high magnetic Reynolds number, which means that magnetic diffusion plays a small part in the transport of the magnetic field, which is mainly achieved through meridional circulation.  Unfortunately, if meridional circulation is the dominant magnetic-field transport mechanism, then the north and south hemispheres can become decoupled, with the dynamo and subsequent sunspot cycle operating independently in each hemisphere (Dikpati and Gilman, 2001; Chatterjee, Nandy, and Choudhuri, 2004).  As observers of surface magnetism, we analyze sunspot data for N--S asymmetries in order to place an upper limit on how out-of-phase the hemispheric sunspot cycles can become.

Solar cycle hemispheric asymmetries, both phase and magnitude, have been observed and documented by many researchers. Temmer \textit{et al.} (2006) report that the solar cycle minima from 1945\,--\,2004 were in phase while the maxima, declining and increasing phases were clearly shifted.   They also found that the N--S asymmetry, based on an absolute N--S asymmetry index, was enhanced at cycle maximum, which is in contradiction with results obtained by Joshi and Joshi (2004) who found the N--S asymmetry was greatest as solar cycle minimum.  Li \textit{et al.} (2002) report that the hemispheres are synchronized during the beginning and end of the cycle but have different maximum amplitudes at different times.    Temmer et al. (2006) tested for the asymmetry of the hemispheres by analyzing the difference between the number of sunspots in the north and south hemispheres.  They were able to confirm a weak magnetic interdependence between the N--S hemispheres proposed by Antonucci \textit{et al.} (1990). In addition, Temmer \textit{et al.} (2002) reported that the northern hemisphere showed evidence of a preferred zone of activity, \textit{i.e.} one or more active longitudes, from the autocorrelations of daily sunspot numbers for Carrington Rotations 1975\,--\,2000, whereas the southern hemisphere did not show evidence for active longitudes.  We embarked on this research with hopes that more extensive analysis of hemispheric activity would better quantify the strength of hemispheric coupling, or find correlations between sunspot cycle characteristics and measures of N--S hemispheric asymmetry.  

Certain sunspot cycle features are indicators of N--S hemispheric asymmetry.  Polar field reversals occuring at different times in the North and South are evidence of hemispheric phase differences.  Durrant and Wilson (2002) reported that the northern hemispheric polar field reversed approximately five months before the southern hemispheric polar field reversal in cycle 23.   In addition, the double peak in sunspot activity is often present in smoothed plots of sunspot number (or sunspot area) as a function in time, and could be caused by the superposition of two hemispheric sunspot indices slightly out of phase.  This feature of the sunspot cycle can be viewed from a different perspective, in that the double peak is not as interesting as the distinctive gap between the peaks when the sunspot numbers diminish for a period of time right during cycle maximum. Known as the Gnevyshev Gap, this was first reported in the literature by Gnevyshev (1963, 1967) for solar cycle $19$, based on observations of the 530.3 nm coronal line.  The Gnevyshev gap was also shown to contain complexity, being not just two peaks and a single gap, but containing multiple peaks and gaps (Kane, 2005).  It was observed that the gap coincided with the period of solar polar heliomagnetic reversal, and was perhaps caused by this global reorganization of solar magnetic fields (Feminella and Storini, 1997; Storini \textit{et al.}, 1997).  Feminella and Storini (1997) developed a method for detecting this gap.  Temmer \textit{et al.} (2006) analyzed sunspot numbers from the north and south hemispheres and found indications that the gap occurs in each hemisphere individually. 

Reports of asynchronous hemispheric behavior and the possibility that the Gnevyshev gap had origins in the averaging of hemispheric indices motivated our research to determine the range of hemispheric phase differences, and in doing so, provide observational constraints for the strength of hemispheric coupling.  To this end, we determine the phase between the hemispheres during the rising, maxima, and declining stages of the solar cycles.  We analyze the frequency of the Gnevyshev gap in hemispheric \textit{versus} total sunspot numbers.  Also, since larger quantities of magnetic flux diffusing across the Equator would cause stronger hemispheric coupling, we search for correlations between observed N--S hemispheric phase lags, length of solar minima, and the measured magnetic flux crossing the Equator.    

\begin{figure}
\includegraphics[width=0.95\linewidth]{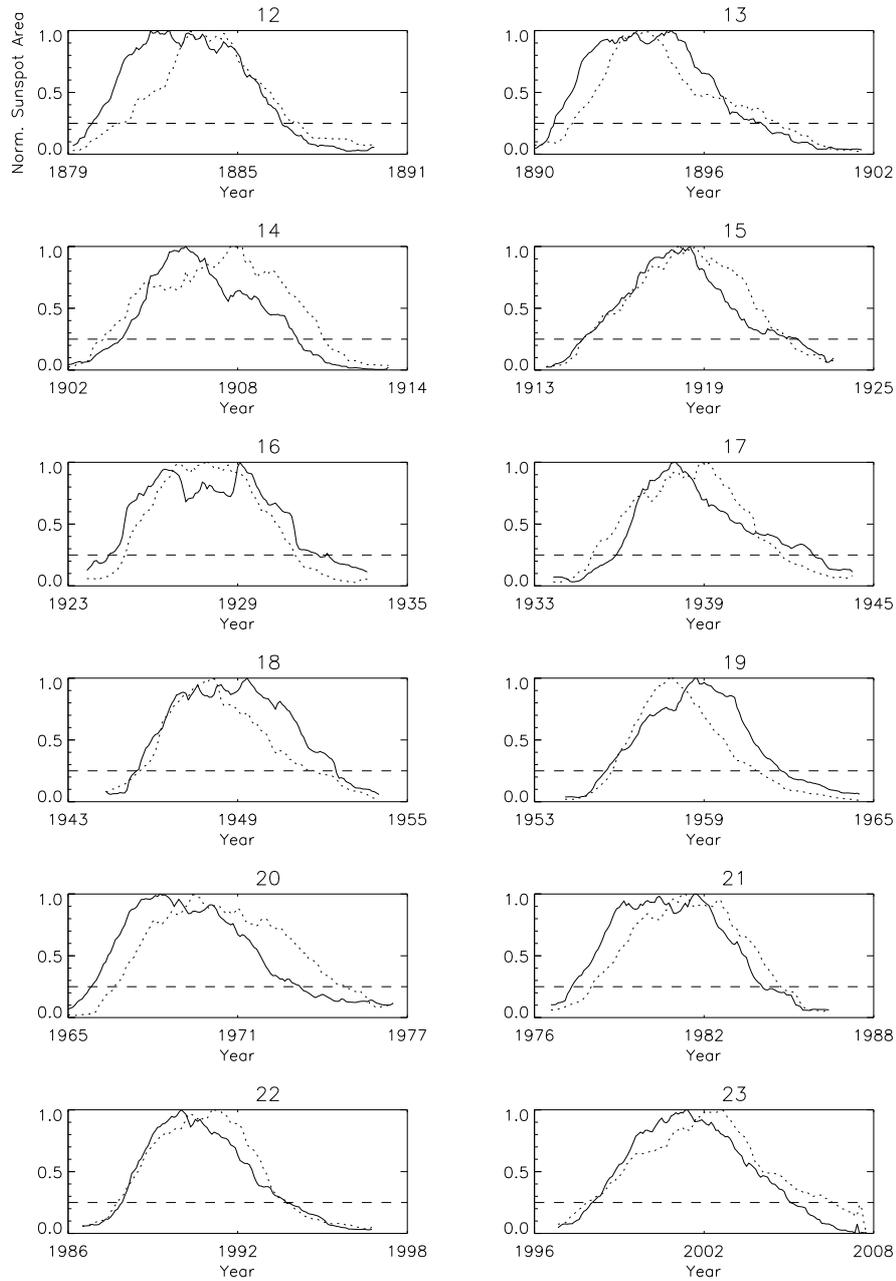}
\caption{Normalized sunspot area smoothed over two years plotted a function of time for solar cycles 12\,--\,23. North (solid) and south (dotted) hemispheric sunspot area data are plotted. A dashed line is drawn at $y=0.25$ to more easily identify the hemispheric phase differences during the rising and declining periods of each solar cycle. \label{fig:1}}
\end{figure}

\section{Data}

%% Greenwich data introduction

Royal Greenwich Observatory (RGO) data retrieved from\\ 
\url{http://solarscience.msfc.nasa.gov/greenwch.shtml} is one of few data archives that provides sunspot data separated into hemispheres. We used the monthly hemispheric sunspot-area values beginning in solar cycle 12 with area in units of millionths of a hemisphere. We analyzed this data to determine phases between the north and south hemispheres.   Another data archive providing hemispheric information is the daily sunspot number from the Kanzelh\"{o}he Solar Observatory (KSO) and Skalnat\'{e} Pleso Observatory (SPO) and reduced by Temmer \textit{et al.} (2006). This data arhive begins in solar cycle 18. The sunspot area and number were used to determine hemispheric phase values and the presence of a Gnevyshev gap.  The Solar Influences Data Analysis Center (SIDC) catalog was also utilized to examine cycle length. We used the SIDC total-sunspot number not separated into hemispheres since this data extends back to 1818, corresponding to the middle of solar cycle 6. The SIDC catalogue does contain sunspot number data separated into hemisphere beginning in 1992 (Vanlommel \textit{et al.}, 2004), but we did not utilize it since the other archival data extended further back in time.   To measure the magnetic flux transport for our research, we used magnetic flux synoptic maps from the Kitt Peak Vacuum Tower (KPVT) and the Synotic Optical Long-term Investigations of the Sun (SOLIS). Carrington rotations 1625\,--\,2007 were from KPVT and 2008\,--\,2071 were from SOLIS. 

\begin{figure}
\includegraphics[width=0.90\linewidth]{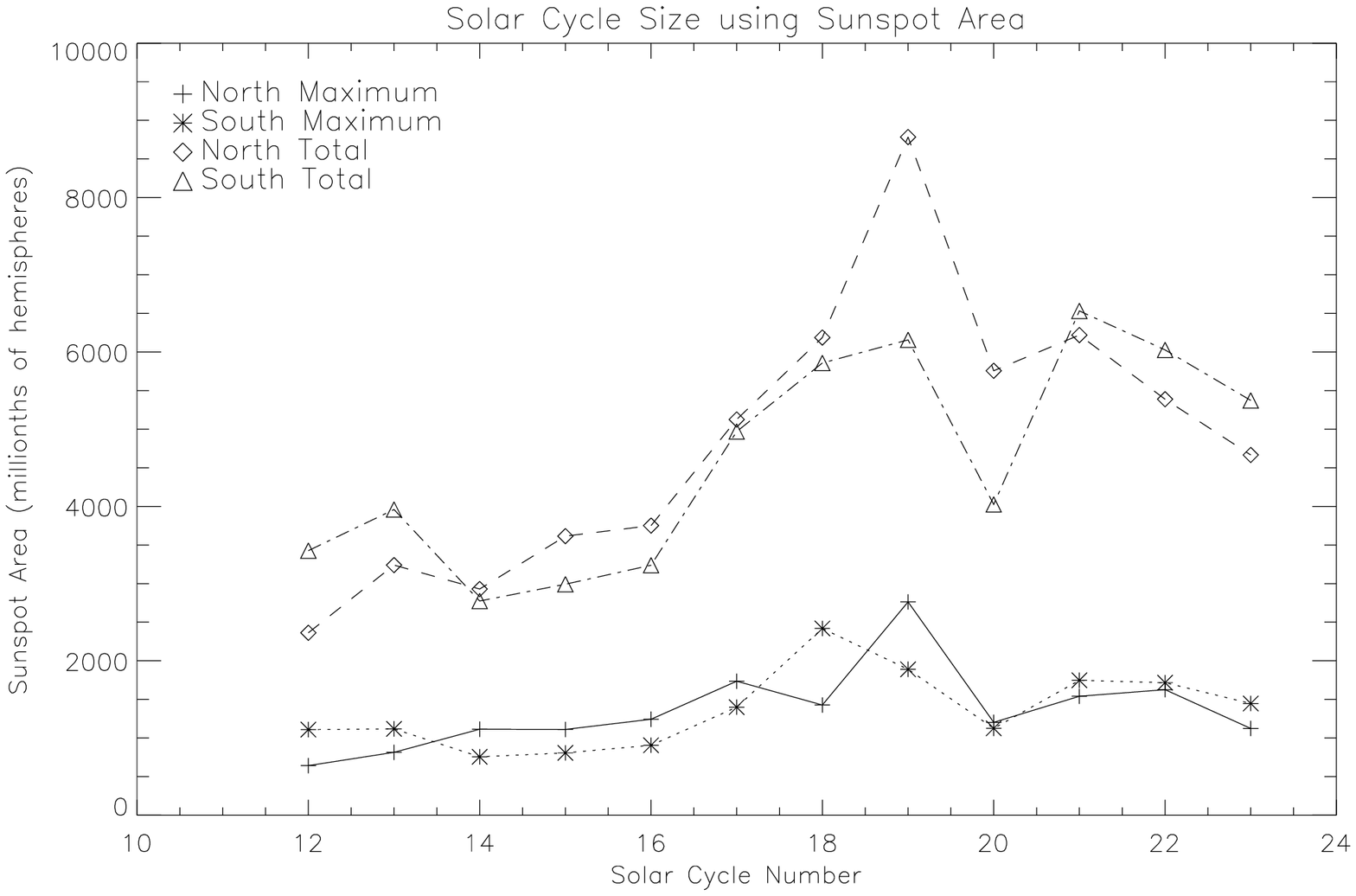}
\includegraphics[width=0.90\linewidth]{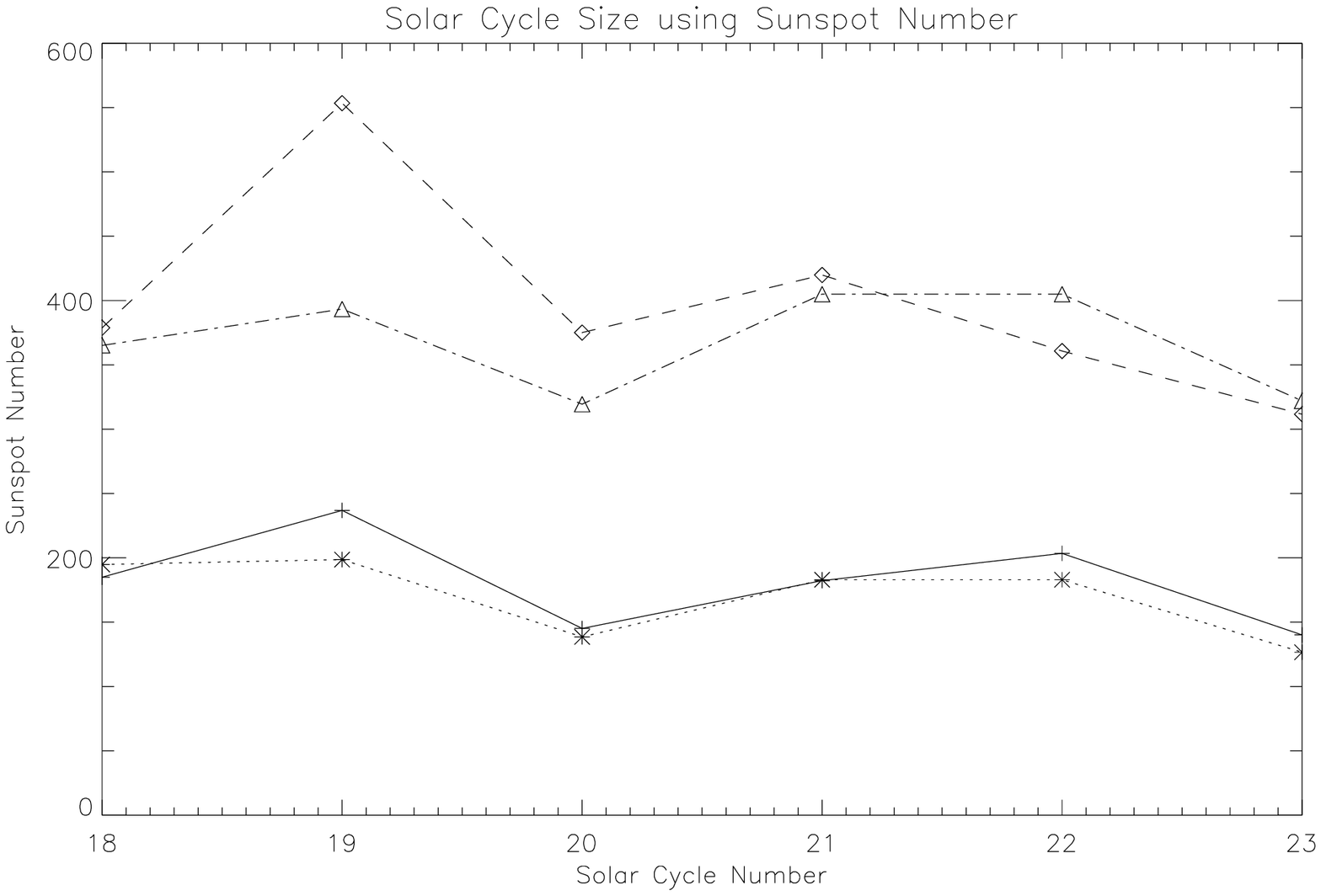}
\caption{Top: The strength, or size, of a given solar cycle is shown as the summed sunspot areas for the north ($\Diamond$) and south ($\triangle$) hemispheres for the duration of each solar cycle, including cycles $12-23$.  The maximum monthly sunspot area value for each solar cycle is also shown for the north (+) and south ($\ast$) hemispheres. Bottom: Same as left using KSO/SPO sunspot number for a shorter period of time including solar cycles $18-23$.\label{fig:2}}
\end{figure}

\section{Separating the Data into Individual Solar Cycles}

To analyze the phase of the solar cycles in each hemisphere, we first had to separate the data into individual solar cycles. Defining a precise date for solar-cycle minimum can be arbitrary and the results vary by several months based on the definition (see Harvey and White, 1999). To isolate each solar cycle, we first smoothed the data eight months for the monthly sunspot area data and 90 days for the daily sunspot number data. This assured less noise at solar-cycle minimum. We then assigned solar cycle minimum as the date that corresponded to the lowest sunspot area or number.  If the date was not unique, we smoothed the data again in order to find a unique minimum, and defined our minimum to be the point of lowest value closest to that distinct minimum. For example, our method resulted in a date of 1996.5 (July 1996) for the solar-cycle minimum between cycles 23 and 24.  Our result compares reasonably well to the cycle-minimum date of 1996.7 (September 1996) as determined by Harvey and White (1999).  Harvey and White used several parameters to determine the date of minimum, including the number of spotless days, number of new- and old-cycle regions as well as a smoothed sunspot number. The time periods resulting from the separation of solar cycles $12-23$ are plotted in Figure 1.  In order to easily characterize the solar cycle strength, we plotted the peak sunspot area and number in each solar cycle as well as total sunspot area in Figure 2.

\section{Identifying the Gnevyshev Gap}

We used the RGO sunspot areas and the KSO and SPO sunspot numbers, to identify the presence of the Gnevyshev gap. We first identified a maximum value in each solar cycle.  Then we determined possible secondary maxima with values at least 50\% of the primary maximum.  It is not a requirement that the secondary maxima occur after the primary maximum. Secondary maxima can be identified anytime within a given solar cycle, including times prior to the primary maximum. If a period of time (at least four months) occurred between the primary and secondary maxima when the values were 25\% lower than the primary maximum, then a Gnevyshev gap was determined to be present. A running four month average was calculated before the gap detection.  Figure 3 shows examples of the gap detection method for total and hemispheric sunspot-area data for cycles 19 and 23. Results for cycles 12$-$23 are shown in Table 1. The Gnevyshev gap is only present in the total sunspot area or number when it is also present in at least one of the hemispheres.  The gap also appears more often in the sunspot-area data than the sunspot number data. 

Note that simply searching for a three $\sigma$ drop persistent over a period of time between the maxima was found to be an unreliable method of identification.  We also tried the method used by Feminella and Storini (1997) who found the gaps were more easily identified if the standard deviation of the monthly mean was plotted with respect to the annual mean. The problem with this method was that the standard deviation is higher at times of rising and falling which caused relative maximum to be detected outside of the solar maxima.

%% Table
%
\begin{table}
\caption{Gnevyshev Gap Detection Results}%\label{tbl:?}
\begin{tabular}{c c c c c c c}
% \begin{tabular}{}     
 \hline
Solar Cycle &
\multicolumn{3}{c} {Sunspot Area} &
\multicolumn{3}{c}{Sunspot Number} \\
\hline
$ $ & Total & North & South & Total & North & South \\
\hline
12 &  & x &  & - & - & - \\
13 &   & x &  & - & - & - \\
14 & x & x & x & - & - & - \\
15 &  &  & x & - & - & - \\
16 & x & x & x & - & - & - \\
17 & x & & x & - & - & - \\
18 & x & x &  & x & x &  \\
19 & & x & &  & x &  \\
20 & x & x & x &  & x & x \\
21 & x & x & x &  & x & x \\
22 & x & x & x &  &  & x \\
23 & x & x & x &  &  & x \\
% <data>
 \hline
 \end{tabular}
 \end{table}

\begin{figure}
\includegraphics[width=0.95\linewidth]{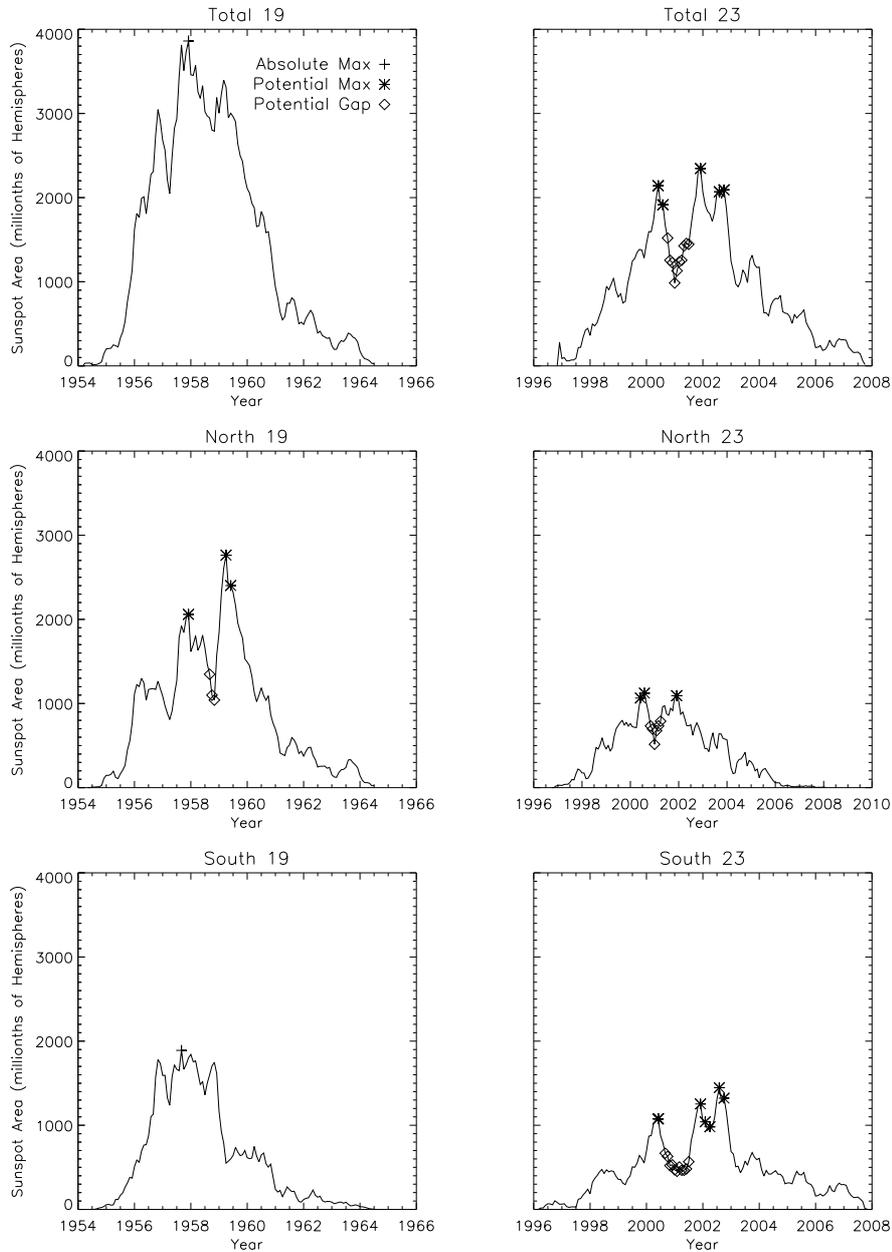}
\caption{An example of detecting the presence of a Gnevyshev Gap for solar cycles 19 and 23 determined from sunspot area data. 
The total sunspot area data (top), the northern hemispheric data (middle), and the southern hemisphere (bottom) are plotted with time periods that could be relative maxima ($\ast$) and Gnevyshev gaps ($\Diamond$) identified.
\label{fig:3}}
\end{figure}

\begin{figure}
\includegraphics[width=0.85\linewidth]{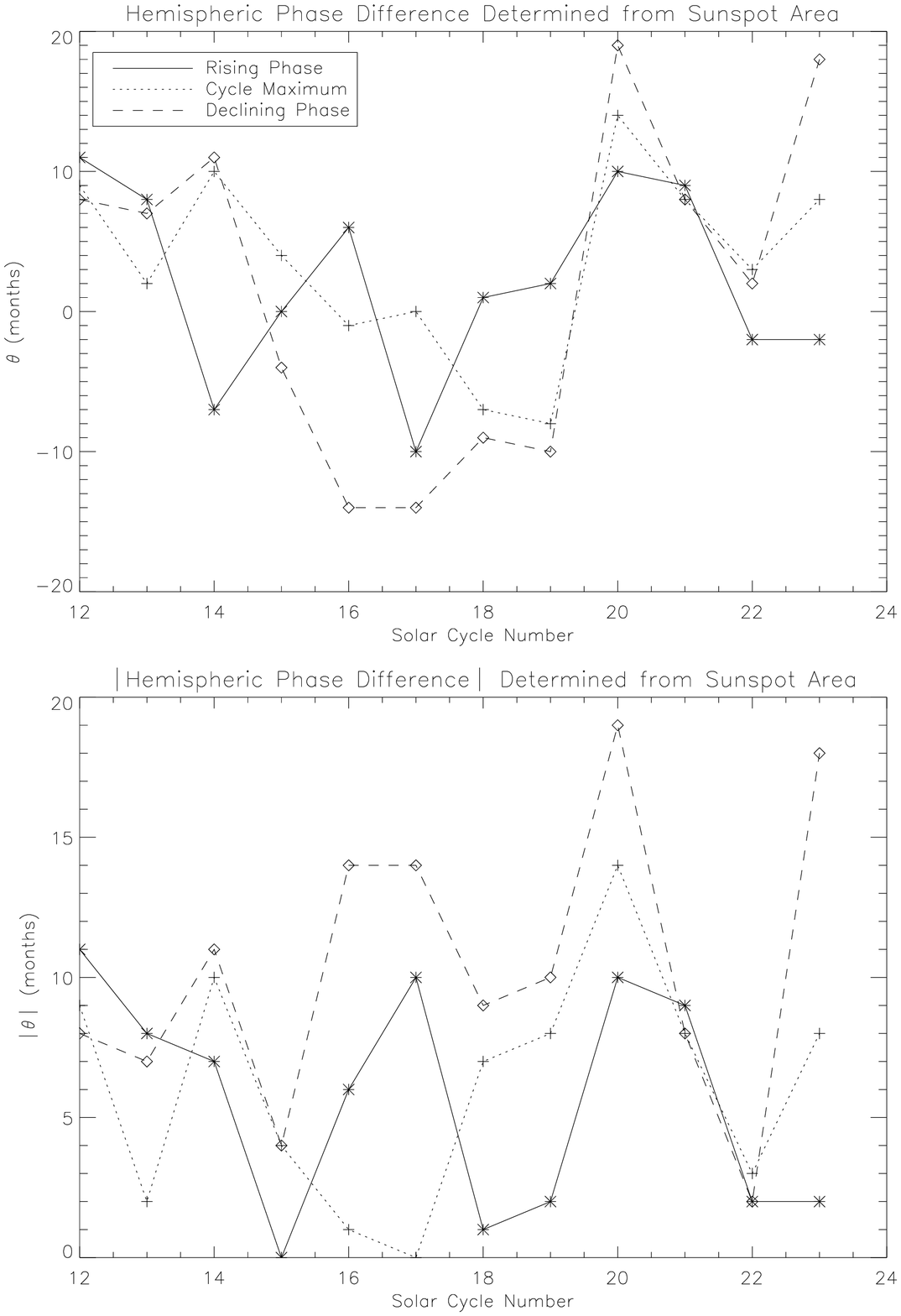}
\caption{Top: Hemispheric phase difference determined from RGO sunspot area for the rising (solid), maximum (dotted), and declining phase (dashed) is shown.  Positive values indicate the North is leading the South.  Phases are determined as the difference in time at which each hemisphere produces 25\% of the maximum value for rising and declining periods of the solar cycle (see Figure 1).  Phases associated with cycle maxima are determined as the differences in time each hemisphere takes to produce half the total sunspot area. Bottom: Absolute value of the phases shown in the top plot.
\label{fig:4}}
\end{figure}

\section{Hemispheric Phase Determination}

In order to determine the phase between the hemispheres during the solar cycle, we explored cross-correlation and onset time. We found cross correlation not optimal because it takes the entire solar cycle into consideration, whereas it is evident that the phasing of the hemispheres can change midway through the cycle. For example, during solar cycle 16 the north hemisphere led the south hemisphere at the beginning of the cycle, but the southern hemisphere led during the declining phase of the solar cycle (see Figure 1). 

Measuring hemispheric phase difference during the rising phase of asolar cycle, or as a solar cycle onset time, was encouraged by Charbonneau (2007) to provide as indication of cross-hemispheric coupling.  After smoothing the data over 24 months and normalizing, we simply identified when the sunspot area reached 25\% of the maximum cycle value in each hemisphere. We repeated this for the declining period of the cycles.  For cycle maxima, we summed the total sunspot area produced during each hemisphere and identified when half the total number of sunspots were produced. Hemispheric differences between these times are reported as the phases in units of months.  Hemispheric phase differences are shown as a function of solar cycle number in Figure 4.  We believe these results are supported qualitatively by the temporal trends seen in Figure 1.  The phase lag determined during cycles 19 and 20, when the South gets ahead of the North, then the trend reverses when the North leads the South by large margins, is consistent with the results of Temmer \textit{et al.} (2006)(see Figure 6 of this paper).

We compared our results for cycle 23 with the polar field reversal. Durrant and Wilson (2002) reported that the polar-field reversal occurred during CR $1975\pm2$ in the north hemisphere and CR $1981\pm1$ in the south hemisphere. This corresponds to a phase difference of $5.4\pm2$ months.  Using RGO sunspot area data, we measured the phase difference to be eight months at the middle of the solar cycle when the polar fields were reversing, whereas the using KSO/SPO sunspot number data determined the phase difference to be six months. Because the RGO sunspot data covers a more extensive time period, we report the hemispheric phase values determined from the RGO data and simply note that the values could be slightly different from KSO/SPO data.

%% Table
%
\begin{table}
\caption{Summary of N--S Hemispheric Phase $|\theta|$ for Cycles $12-23$ (units are in months except for kurtosis which is unitless)}%\label{tbl:2}
\begin{tabular}{ c c c c c c c }
\hline
$ Cycle Period$ & Average Total Time & $|\theta|_{max}$ & $|\theta|_{avg}$ & $\sigma$ & Kurtosis \\
\hline
Rising    & 21 & 11& 5.7& 4.0 & -1.8 \\
Declining & 40 & 19& 10.3&5.2 & -1.1\\
% <data>
 \hline
 \end{tabular}
 \end{table}

The hemispheres may be more strongly coupled at solar cycle minimum, due to magnetic-flux cancellation across the Equator when the toroidal bands are closer to the equator, than at solar-cycle maximum when the toroidal bands are at higher latitudes. We examined whether hemispheric-phase differences were significantly larger during the rising or declining phase of the cycle.  Solar cycles are not perfectly symmetrical in rising and declining phases. Most solar cycles rise faster than they fall.

The rising period of the solar cycle is $21$ months on average during solar cycles $12-23$.  The maximum and mean observed values of the N--S hemispheric phase lag during the rising period are $11$ and $5.7$ respectively. The standard deviation of the mean value is $4$ months and the data has a negative kurtosis value.  Kurtosis is a measure of the ``peakedness'' of the distribution.  A negative kurtosis indicates that phase values in the rising stage of the solar cycle are not distributed around a single peak (not Gaussian or normally distributed). The histogram (not plotted) shows the phase values to be randomly distributed across the range, although 12 cycles provides only small-number statistics.  The maximum phase difference is approximately half of the solar cycle rising time.  See Table 2 for a summary of the values.  

The declining period of the solar cycle is 40 months long on average during solar cycles $12-23$.  The maximum and mean observed values of the hemispheric phase lag during the rising period are $19$ and $10.3$ respectively. The standard deviation of the mean value is $5.5$ months and the data has a negative kurtosis value.  Again, the distribution of hemispheric phases in the declining stage of the solar cycle appears random and not distributed around a single peak.  Again, the maximum phase difference is approximately half of the solar cycle declining time.   See Table 2 for a summary of the values.

If the hemispheres were greatly out of phase at the end of a given solar cycle, it is possible a lengthier minimum may result to allow time for hemispheric coupling to take place. To test this, we plotted phase difference \textit{versus} length of solar cycle, see Figure 5. The length of solar cycle minimum was defined as the period of time when the sunspot values first decreased below 20\% of the previous maximum value until they first exceeded 20\% of the next maximum value. We found no correlation between length of minimum and phase, see Figure 5 bottom panel.

\begin{figure}
\includegraphics[width=0.85\linewidth]{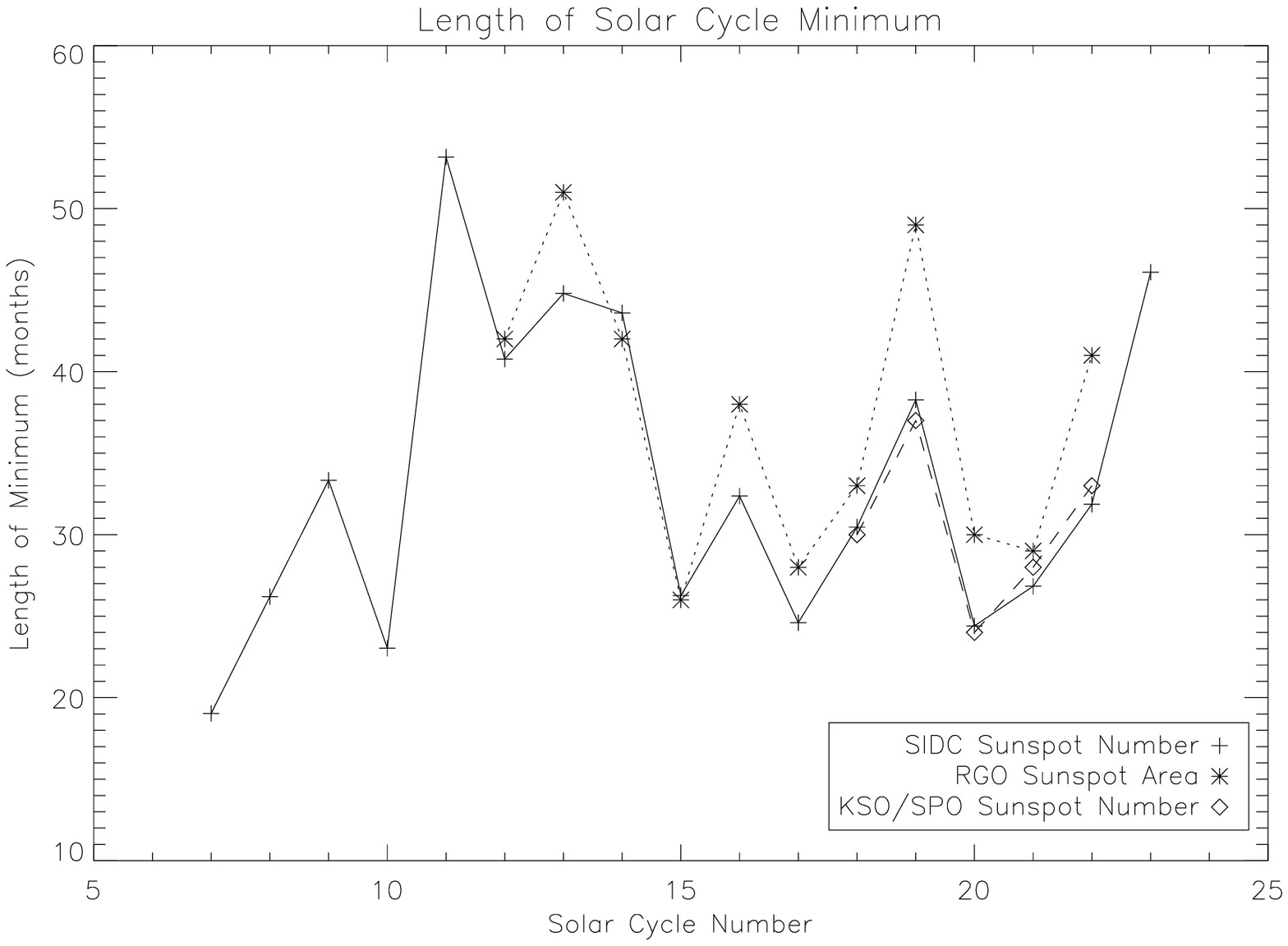}
\includegraphics[width=0.85\linewidth]{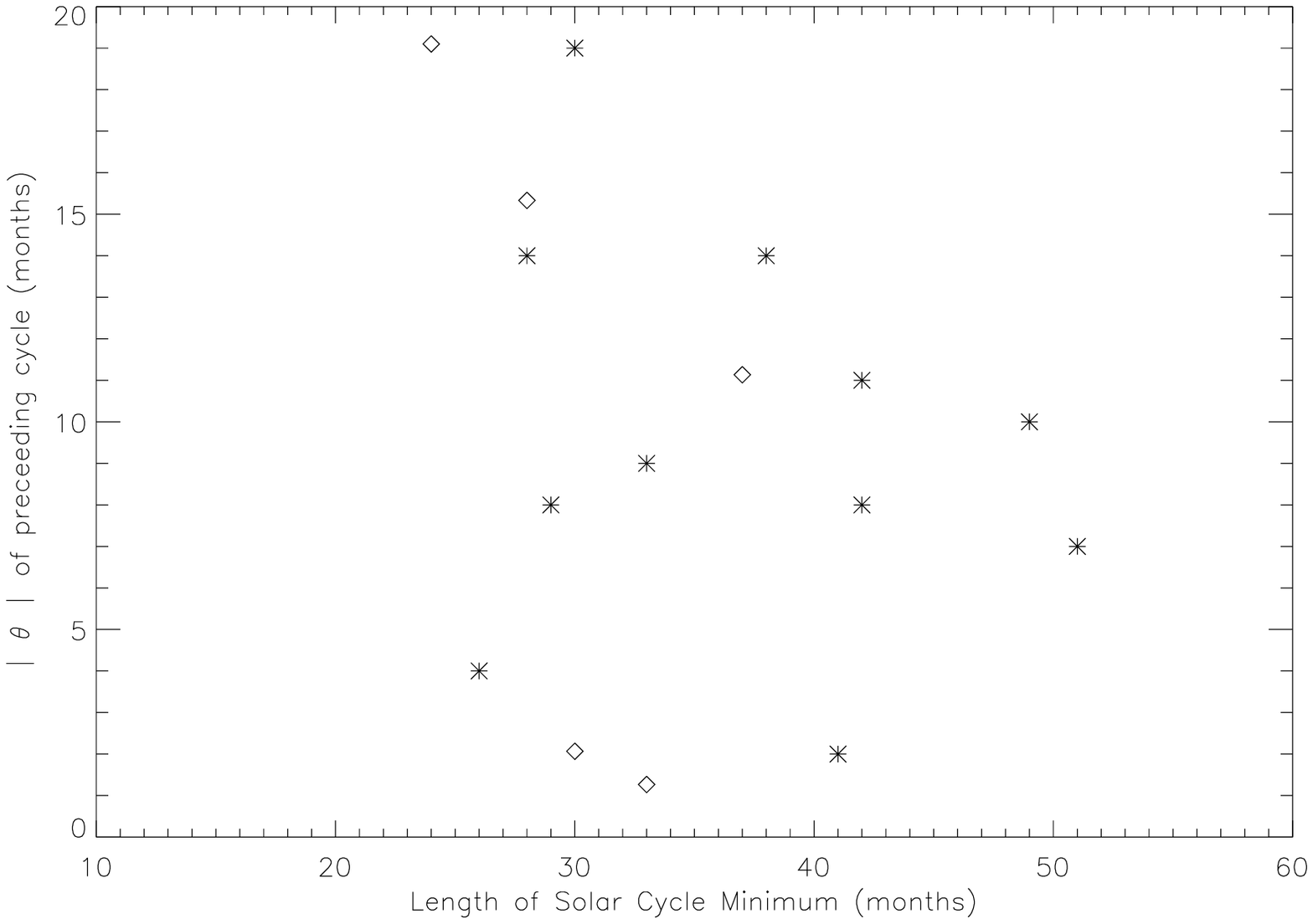}
\caption{Top: The length of solar-cycle minimum as determined from the RGO sunspot area data ($\ast$), the KSO/SPO sunspot number data ($\Diamond$), and the SIDC sunspot number data (+) is plotted as a function of the solar cycle number just preceeding the minimum. The SIDC and KSO/SPO data both show the same trends. Solar Cycle 23 minimum length is only determined from the SIDC sunspot number since it is the most up to date (through the end of August 2009). The minimum length is defined as the time period between cycles when the preceding cycle is at least 20\% of its maximum value until the next cycle reaches 20\% of its maximum value. Bottom: The hemispheric phase of the preceeding solar cycle (determined during the declining period of that solar cycle) is plotted as a function of the length of solar cycle minimum for RGO ($\ast$) and KSO/SDP sunspot data ($\Diamond$).  There is no obvious correlation between length of minimum and the hemispheric phase.  \label{fig:5}}
\end{figure}
\section{Examining the Minimum Between Cycles 23 and 24}

The minimum after cycle 23 is not included in Figure 5 (except as SIDC data for length of minimum) and in the discussion in Section 5 because it is not complete. We have, however, analyzed some aspects of it to date. This minimum has seemed longer and quieter than most. Only a few sunspots from cycle 24 have appeared.  The average length of solar-cycle minima as seen in Figure 5 is $32, 37$, and $30$ months using SIDC, RGO, and KSO/SPO data, respectively. Cycle 23 minimum is not included in these mean calculations. The cycle 23 minimum is 46 months in length to date (31 August 2009) as calculated by the SIDC sunspot number data. It is still close to average in length from a historical perspective, but certainly more similar to the minima experienced in the late 1800s and early 1900s as opposed to the more recent minima.   

This minimum has also had a large number of spotless days, so we examine the number  of spotless days in each solar cycle. Figure 6 shows that the number of spotless days is not significantly higher during the cycle 23 minimum from a historical perspective, but it is higher than any minimum that we have experienced since the the minimum between cycle 14 and 15 in 1913.   

\begin{figure}
\includegraphics[width=0.85\linewidth]{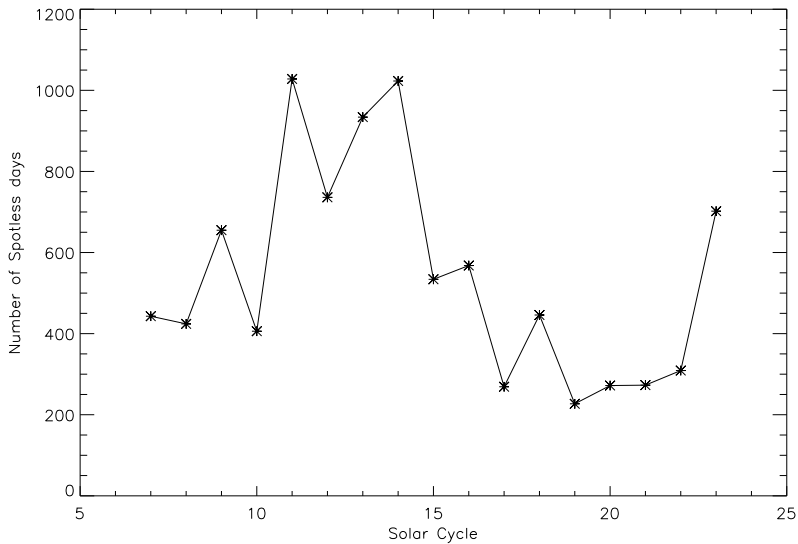}
\caption{The number of spotless days in solar cycle minimum according to SIDC sunspot number. The solar-cycle number corresponds to the solar cycle preceding the solar cycle minimum. 
Data set ends 31 August 2009. \label{fig:6} }
\end{figure}

\section{Magnetic Flux Transport}

Magnetic flux transport is critical to understanding the nature of the solar dynamo.  In particular, flux crossing the Equator by diffusion can provide a direct measure of cross-hemispheric coupling. According to Durrant \textit{et al.} (2004) the flux transport equation at a specific latitude and longitude on the solar surface is:
\begin{equation}
\frac{\delta \Phi}{ \delta t}(\theta, \phi)=\frac{\pi R_{\odot} cos(\theta)}{180}
\left(
-v_{mf}(\theta)B(\theta, \phi)+\frac{\eta}{R_{\odot}}\frac{\delta B(\theta, \phi)}{\delta \theta} 
\right),
\end{equation}
where $B$ is the magnetic field, $v_{mf}$ is the meridional-flow velocity, and $\theta$ and $\phi$ are the latitude and longitude respectively. At the Equator, the velocity of meridional flow is on average zero.  Equation 1 then becomes:
\begin{equation}
\frac{\delta \Phi}{ \delta t}(0, \phi)=\frac{\eta \pi }{180}  \frac{ \delta B}{\delta \theta}(0, \phi).
\end{equation}
We determined the magnetic flux crossing the Equator as a function of time using KPVT maps for Carrington Rotations $1625-2007$ and SOLIS Carrington Rotation maps from $2008-2071$.  
%The flux crossing the equator was determined by averaging 3 equal increments in sine latitude in each direction from the Equator.  
Our results are shown in Figure 7.  Our results confirm Cameron and Sch\"{u}ssler(2007) who show that the magnetic flux crossing the equator reaches a maximum slightly after the peak of the solar cycle. 

There could be a correlation between the amount of the magnetic flux crossing the Equator and the magnitude of the phase difference between the hemispheres.  To test this, we determined the hemispheric phase difference as a function of the flux crossing the Equator of preceding cycle and found no correlation. Therefore, our examinations of surface flux crossing the equator provided no insight into the understanding of a physical mechanism that limited hemispheric phase differences.  However, it is possible that there is un-observed cross-hemispheric coupling in the form of magnetic-flux transport or magnetic reconnection occuring in near-surface layers or at depths closer to the tachocline.

\begin{figure}
\includegraphics[width=0.85\linewidth]{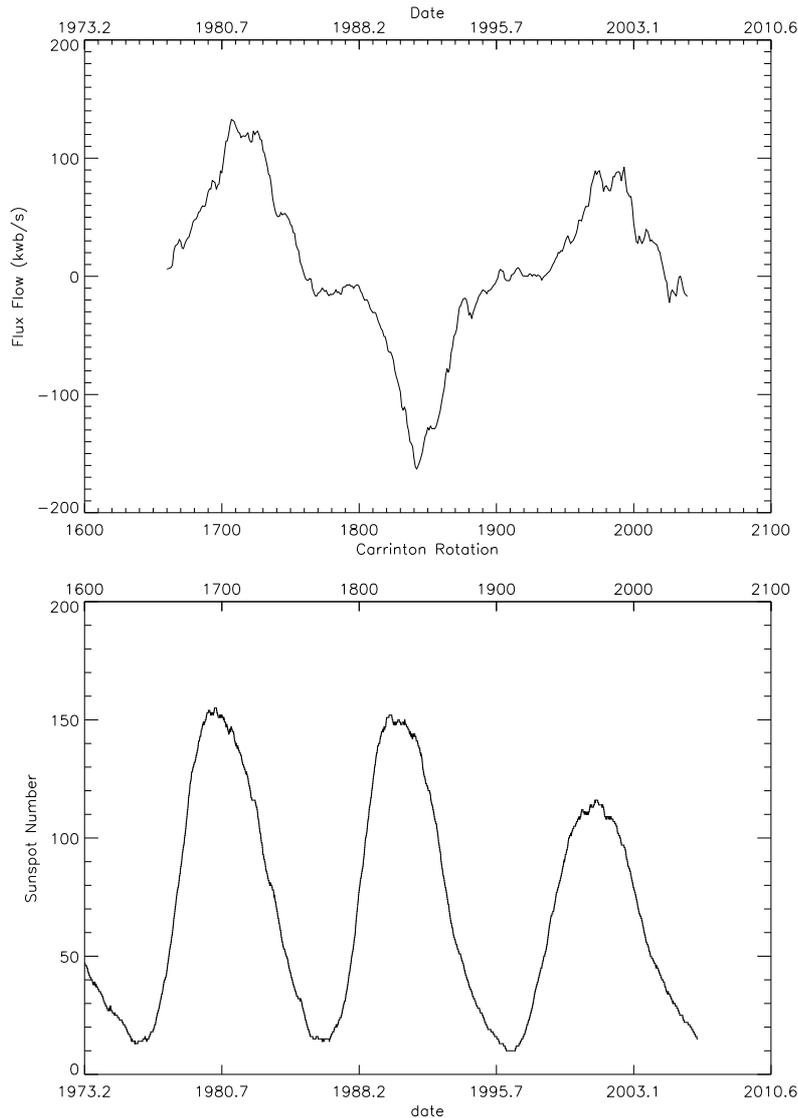}
\caption{A smoothed measure of magnetic flux crossing the Equator as determined using the Durrant \textit{et al.} (2004) methodology is plotted as a function of Carrington Rotations $1625-2071$. Positive values indicate north magnetic polarity transferring to the southern hemisphere. The SIDC sunspot number, smoothed by calculating a running average over a 30-month period, is plotted below. 
\label{fig:7}}
\end{figure}

\section{Conclusions}

Though the hemispheres are often observed to be temporally out of phase during any given solar cycle, we found that they are never more than a few years out of phase.  The hemispheric phase difference was most consistently determined by smoothing and normalizing the hemispheric sunspot area data, then defining the phase during the rising and declining period of the cycle as when each hemisphere reached 25\% of its maximum value. We found the maximum and mean phase difference to be $11$ and $5.7$ months in the rising phase of the solar cycle and $19$ and $10.3$ months in the declining phase of the solar cycle. The maximum observed phase lag was roughly half of the total duration of the rising and declining periods (21 and 40 months, respectively).  The phase values appear to be randomly distributed between 0 months (in phase) and the maximum values.   Based on our sunspot analysis from cycles $12-23$, we conclude that dynamo models that produce sunspot cycles with the hemispheres more than $11$ ($19$) months out of phase during the rising (declining) periods of the sunspot cycle are not supported by observations.  Instead, the temporal phase lags between the hemispheres should be randomly distributed between a zero phase lag and the maximum value of half the time that it takes for the solar cycle to rise and/or decline.  

One possible explanation for a randomly distributed hemispheric phase difference could be found in Ossendrijver and Hoyng's (1996) examination of the variations in cycle amplitude and phase caused by rapid, stochastic variations in the kinetic helicity of a plane-parallel dynamo model.  The fluctuations are modeled as a random forcing term that could be operating independently in the hemispheres.  We mention this, but emphasize that it is outside the scope of this paper to comment on adopted parameter values in dynamo models, including the varied magnetic-diffusivity values of mean-field and Babcock--Leighton models that produce too strong and too weak hemispheric coupling. We only attempt to provide an observational measure of the temporal phase lag of the hemispheres during cycles $12-23$.  It is obvious that some degree of cross hemispheric coupling prevents the hemispheres from becoming grossly out of phase and sets an upper limit to the extent to which they may be out of phase (see Charbonneau, 2005, 2007). 

The rest of our analysis either confirmed previous findings or were null results.   We found no correlation between hemispheric phase difference and length of the solar cycle.  We found no correlation between the amount of magnetic flux crossing the Equator during a solar cycle and the hemispheric phase difference.   We find that 8 out of 12 solar cycles have a detectable Gnevyshev gap in the total sunspot data, where 10 (8) out of 12 have a detectable gap in the northern (southern) hemispheric sunspot data, see Table 1. A gap was never apparent in the total sunspot record when it was not apparent in at least one of the hemispheric records.  We conclude that the double-peaked sunspot cycle and the resulting Gnevyshev gap is not an artifact caused by averaging data from hemispheres out of phase.  Rather, we confirm previous results that it is implicit in each hemisphere and must be caused by a physical mechanism operating within each hemisphere. 

As an aside, our examination of the current cycle 23 minimum, as seen in Figures 5 and 6, shows it to not be exceptionally quiet in an historically sense, although it is certainly more comparable to the minima experienced in the late 1800s and early 1900s as opposed to more recent minima. However, this minimum is still not complete.  It has produced the greatest number of spotless days (as of 31 August 2009) since the minimum between cycle 14 and 15 in 1913.

%%%%%%%%%%%%%%%%%%%%%%%%%%%%%%%%%%%%%%%%%%%%%%%%%%%%%%%%%%%%%%%%%%%%%%%%%%%
%% Acknowledgements
%
\begin{acks}
This work was carried out through the National Solar Observatory Research Experiences for Undergraduate (REU) site program, which is co-funded by the Department of Defense in partnership with the National Science Foundation REU Program.  SOLIS data used here are produced cooperatively by NSF/NSO and NASA/LWS.
\end{acks}

%%% %%%%%%%%%%%%%%%%%%%%%%%%%%%%%%%%%%%%%%%%%%%%%%%%%%%%%%%%%%%
%% Bibliography

\end{article} 
\end{document}